\documentclass[traditabstract]{aa}
\usepackage{graphics}
\usepackage{graphicx}
\usepackage{multicol}
\usepackage[varg]{txfonts}

\usepackage{natbib,twoopt}
\usepackage{amsmath} 
\usepackage[breaklinks=true]{hyperref} 
\bibpunct{(}{)}{;}{a}{}{,} 
\makeatletter
\newcommandtwoopt{\citeads}[3][][]{\href{http://adsabs.harvard.edu/abs/#3}%
{\def\hyper@linkstart##1##2{}%
\let\hyper@linkend\@empty\citealp[#1][#2]{#3}}}
\newcommandtwoopt{\citepads}[3][][]{\href{http://adsabs.harvard.edu/abs/#3}%
{\def\hyper@linkstart##1##2{}%
\let\hyper@linkend\@empty\citep[#1][#2]{#3}}}
\newcommandtwoopt{\citetads}[3][][]{\href{http://adsabs.harvard.edu/abs/#3}%
{\def\hyper@linkstart##1##2{}%
\let\hyper@linkend\@empty\citet[#1][#2]{#3}}}
\newcommandtwoopt{\citeyearads}[3][][]%
{\href{http://adsabs.harvard.edu/abs/#3}
{\def\hyper@linkstart##1##2{}%
\let\hyper@linkend\@empty\citeyear[#1][#2]{#3}}}
\makeatother

\usepackage{color}
\definecolor{mygreen}{RGB}{0,128,0}

\hypersetup{colorlinks=true,linkcolor=blue,citecolor=blue,urlcolor=blue}

\begin{document}

\title{The close circumstellar environment of Betelgeuse\thanks{Based on observations made with ESO telescopes at Paranal Observatory, under ESO program 095.D-0309(E).}}
\subtitle{III. SPHERE/ZIMPOL imaging polarimetry in the visible}
\titlerunning{The visible close-in envelope and photosphere of Betelgeuse}
\authorrunning{P. Kervella et al.}
%
\author{
P.~Kervella\inst{1,2}
\and
E.~Lagadec\inst{3}
\and
M. Montarg\`es\inst{4,2}
\and
S.~T.~Ridgway\inst{5}
\and
A.~Chiavassa\inst{3}
\and
X.~Haubois\inst{6}
\and
H.-M.~Schmid\inst{7}
\and
M.~Langlois\inst{8}
\and
A.~Gallenne\inst{9}
\and
G.~Perrin\inst{2}
}
\institute{
Unidad Mixta Internacional Franco-Chilena de Astronom\'{i}a (UMI 3386), CNRS/INSU, France
\& Departamento de Astronom\'{i}a, Universidad de Chile, Camino El Observatorio 1515, Las Condes, Santiago, Chile, \email{pkervell@das.uchile.cl}.
\and
LESIA (UMR 8109), Observatoire de Paris, PSL, CNRS, UPMC, Univ. Paris-Diderot, 5 place Jules Janssen, 92195 Meudon, France, \email{pierre.kervella@obspm.fr}.
\and
Laboratoire Lagrange (UMR 7293), UNSA, CNRS, Obs. de la C\^ote d'Azur, Bd de l'Observatoire, F-06304 Nice cedex 4, France.
\and
Institut de Radio-Astronomie Millim\'etrique, 300 rue de la Piscine, 38406 St Martin d'H\`eres, France.
\and
National Optical Astronomy Observatories, 950 North Cherry Avenue, Tucson, AZ 85719, USA.
\and
European Southern Observatory, Alonso de C{\'o}rdova 3107, Casilla 19001, Santiago 19, Chile.
\and
Institute for Astronomy, ETH Zurich, 8093 Zurich, Switzerland.
\and
Obs. de Lyon, CRAL, ENS Lyon, CNRS, Univ. Lyon 1, UMR 5574, 9 avenue Charles Andr\'e, 69230 Saint-Genis Laval, France.
\and
Universidad de Concepci{\'o}n, Departamento de Astronom\'{\i}a, Casilla 160-C, Concepci{\'o}n, Chile.
}
\date{Received ; Accepted}
\abstract
{
The physical mechanism through which the outgoing material of massive red supergiants is accelerated above the escape velocity is unclear.
Thanks to the transparency of its circumstellar envelope, the nearby red supergiant Betelgeuse gives the opportunity to probe the innermost layers of the envelope of a typical red supergiant down to the photosphere, i.e. where the acceleration of the wind is expected to occur.
We took advantage of the SPHERE/ZIMPOL adaptive optics imaging polarimeter to resolve the visible photosphere and close envelope of Betelgeuse.
We detect an asymmetric gaseous envelope inside a radius of 2 to 3 times the near-infrared photospheric radius of the star ($R_\star$), and a significant H$\alpha$ emission mostly contained within 3\,$R_\star$. 
From the polarimetric signal, we also identify the signature of dust scattering in an asymmetric and incomplete dust shell located at a similar radius.
The presence of dust so close to the star may have a significant impact on the wind acceleration through radiative pressure on the grains.
The 3\,$R_\star$ radius emerges as a major interface between the hot gaseous and dusty envelopes.
The detected asymmetries strengthen previous indications that the mass loss of Betelgeuse is likely tied to the vigorous convective motions in its atmosphere.
}

\maketitle

\newpage


\section{Introduction}

The role of convection in triggering the mass loss of evolved supergiants is still largely unclear. As the closest red supergiant (RSG), \object{Betelgeuse} is the best object to study the photosphere and close circumstellar envelope (CSE), thanks to its large limb darkened disk angular diameter ($\theta_\mathrm{LD}=42.3 \pm 0.4$\,mas in the near-infared, from \citeads{2014A&A...572A..17M}) and relatively transparent CSE. An overview of recent works on this star is presented in \citetads{2013EAS....60.....K}.
From spectroscopic observations, \citetads{2007A&A...469..671J} proposed that convection, by lowering the effective gravity, could levitate the material above the photosphere and trigger the outflow in conjunction with radiative pressure on molecular lines. Large convective cells of the RSGs were predicted by \citetads{1975ApJ...195..137S} and imaged on Betelgeuse using interferometry by \citetads{2009A&A...508..923H} (see also \citeads{2010A&A...515A..12C}).
\citetads{2011A&A...529A.163O} reconstructed 1D profiles from spectrally dispersed interferometry, and identified upwards and downwards motions of CO that point to the role of convection in triggering the mass loss.
Such behavior is consistent with the plumes extending to a few stellar radii in the near-infrared \citepads{2009A&A...504..115K}.
\citetads{2010A&A...516L...2A} discovered a magnetic field in Betelgeuse from spectro-polarimetry, a factor that could  contribute to enhancing the outflow.
Thermal emission from dust and molecules has also been detected close to the star by \citetads{2007A&A...474..599P}.
The most important region where the wind is accelerated is located within a few stellar radii of the photosphere \citepads{2009ApJ...701.1464H}, i.e. at angular separations $\lesssim 0.1\arcsec$.
The new Spectro-Polarimetric High-contrast Exoplanet REsearch (SPHERE) instrument provides high Strehl adaptive optics (AO) correction in the visible (Strehl ratio $S \approx 30$\%, Table~\ref{obs_log}). This gives access to a resolution of $\approx 20$\,milliarcseconds (mas), thus resolving the photosphere and close-in CSE of Betelgeuse. The other AO systems currently operating at visible (or near-visible) wavelengths are MagAO ($\lambda \gtrsim 600$\,nm, \citeads{2013ApJ...774...94C}) and GPI ($\lambda \gtrsim 900$\,nm, \citeads{2015ApJ...799..182P}).

\section{Observations and data reduction}\label{observations}

SPHERE \citepads{2008SPIE.7014E..18B, 2014SPIE.9148E..1UF} is a high-performance AO installed at the Nasmyth focus of Unit Telescope~3 of the Very Large Telescope. Our observations were obtained on the night of 30 March 2015 using the Zurich IMaging POLarimeter (ZIMPOL, \citeads{2014SPIE.9147E..3WR}).
We observed \object{Betelgeuse} and a point spread function (PSF) calibrator, $\phi^2$\,Ori (\object{HD 37160}, spectral type K0IIIb) in the polarimetric P2 mode. $\phi^2$\,Ori is unresolved by ZIMPOL ($\lambda/D \approx 14-19$\,mas, while $\theta_\mathrm{LD}=2.20 \pm 0.02$\,mas, \citeads{2002A&A...393..183B}), and is also a photometric calibrator \citepads{1999AJ....117.1864C}.
ZIMPOL includes an imaging polarimeter based on the fast modulation technique, with two cameras (hereafter {\tt cam1} and {\tt cam2}) that allow for simultaneous observations at two different wavelengths \citepads{2012SPIE.8446E..8YS}.
We observed Betelgeuse and the PSF reference using four filters: $V$, CntH$\alpha$, NH$\alpha$ and TiO717.
The observations (Table~\ref{obs_log}) required 1.5\,hours of telescope time, and were taken without coronagraph. The total integration time for the two cameras is 1.3\,hours. Despite the extreme brightness of Betelgeuse, a neutral density of only ND1 was necessary in the CntH$\alpha$ and NH$\alpha$.
In the $V$ and Ti0717 filters, we used a ND2 filter. The transmissions are 8.2\% for the ND1 (uniformly for all filters) and between 0.62\% and 0.89\% for ND2 (SPHERE User Manual P95.2). A total of 48 data cubes were recorded on Betelgeuse (10 or 30 images per cube) and 52 cubes on $\phi^2$\,Ori (6 images per cube).
We processed the raw cubes using the data reduction pipeline (v. 0.14.0) and custom {\tt python} routines following the procedure described by \citetads{2015A&A...578A..77K}. We derive the total intensity $I$, the polarized flux $P$, the degree of linear polarization $p_L$, and the polarization electric-vector position angle $\theta$.
\begin{figure}[]
        \centering
        \includegraphics[width=4.4cm]{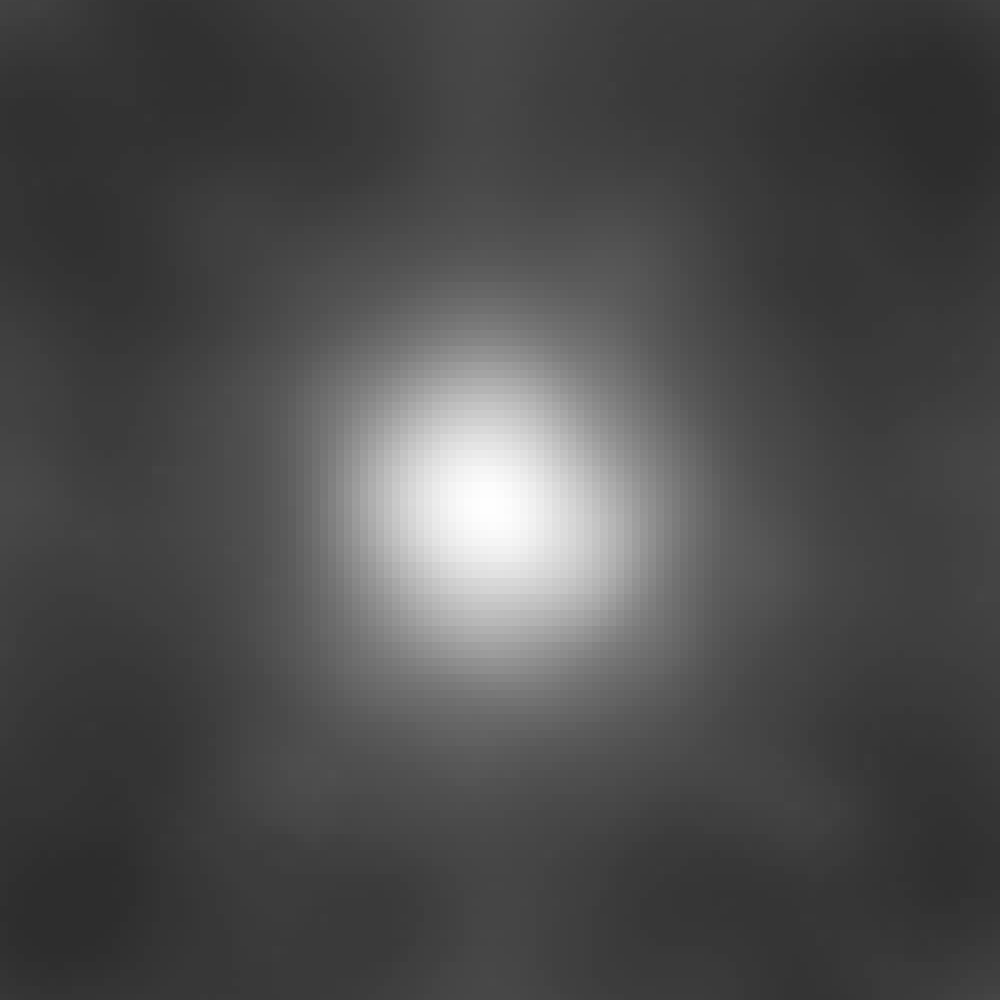}
        \includegraphics[width=4.4cm]{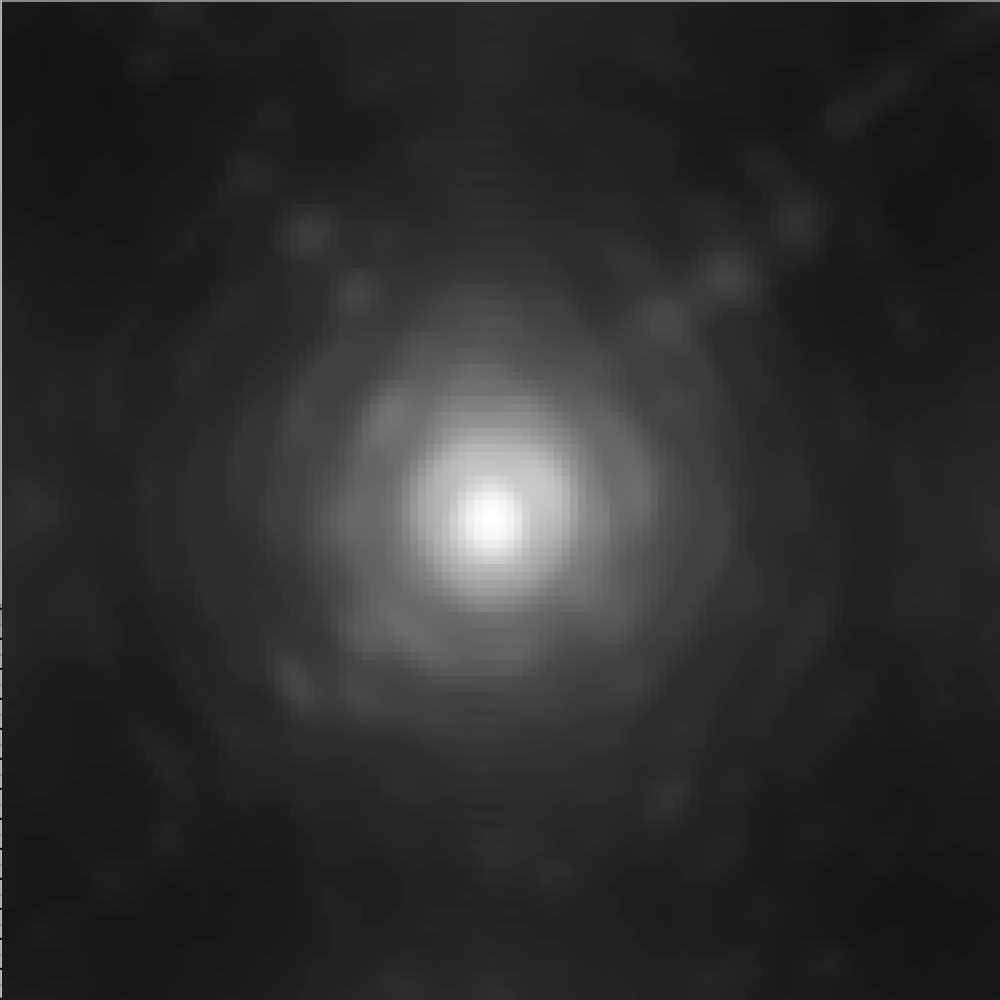}
        \caption{{\it Left:} Non-deconvolved intensity image of Betelgeuse (left) and the PSF reference $\phi^2$\,Ori (right) in the CntH$\alpha$ filter. The gray scale is logarithmic and the field of view is $0.45\arcsec \times 0.45\arcsec$. 
        \label{nondeconv}}
\end{figure}
The plate scale is $3.628 \pm 0.036$\,mas\,pix$^{-1}$ and the position angle of the vertical axis with respect to north is $357.95 \pm 0.55\,\deg$ for both {\tt cam1} and {\tt cam2} (Ch. Ginski, private communication).
Examples of the resulting images of Betelgeuse and the PSF calibrator are presented in Fig.~\ref{nondeconv}.
The full width at half maximum $\theta_\mathrm{PSF}$ of the PSF star images is around 20\,mas in all filters, as listed in Table~\ref{obs_log}. Betelgeuse is thus resolved with approximately two resolution elements over its angular diameter.
We based our flux calibration of the intensity frames of Betelgeuse on those of $\phi^2$\,Ori, using photometry over a large $1.8\arcsec$ circular aperture encompassing the AO halo. The resulting calibrated fluxes measured on Betelgeuse are listed in Table~\ref{phot}.
We estimate the uncertainty on these fluxes to $\pm 5\%$. The magnitudes of Betelgeuse are found to be $m_V = 0.17 \pm 0.05$ and $m_R=-1.07 \pm 0.05$ (TiO717 filter). This is brighter than its average magnitude, consistent with the AAVSO light curve that shows a brightening of $0.3$\,mag between January and April 2015.

\begin{table*}
        \caption{Log of the SPHERE/ZIMPOL observations of Betelgeuse and $\phi^2$\,Ori (PSF calibrator).}
        \centering          
        \label{obs_log}
        \begin{tabular}{clccllccccccccc}
	\hline\hline
        \noalign{\smallskip}
        No  & Star & MJD & Cam. & Filter & ND & Trans. & DIT [s] & Dither & $\Sigma$DIT & $\theta$ & AM & $S$ & $\theta_\mathrm{PSF}$ & PA \\
          & & -57110 & & & & [$\%$] &  $\times$ NDIT & $\times$\,Exp.[s] & [s] & [''] & & [$\%$] & [mas] & [deg]\\
        \hline               
        \noalign{\smallskip}
        1 & $\alpha$\,Ori & 1.9965 & {\tt cam1} & TiO717 & ND2 & 0.89 & $ 3.5 \times 10 $ & $3 \times 2$ & 840 & 0.93 & 1.34 & & $-$ & \\
         & $\alpha$\,Ori & 1.9965 & {\tt cam2} & $V$ & ND2 & 0.62 & $ 3.5 \times 10 $ & $3 \times 2$ & 264 & 0.93 & 1.34 & & $-$ & \\
        \hline                      
        \noalign{\smallskip}
        2 & $\alpha$\,Ori & 2.0096 & {\tt cam1} & CntH$\alpha$ & ND1 & 8.20 & $ 1.1 \times 30 $ & $3 \times 2$ & 792 & 0.88 & 1.41 & & $-$ & \\
         & $\alpha$\,Ori & 2.0096 & {\tt cam2} & NH$\alpha$ & ND1 & 8.20 & $ 1.1 \times 30 $ & $3 \times 2$ & 792 & 0.88 & 1.41 & & $-$ & \\
        \hline                      
        \noalign{\smallskip}
        3 & $\phi^2$\,Ori & 2.0260 & {\tt cam1} & TiO717 & ND1 & 8.20 & $ 3.5 \times 6 $ & $3 \times 2$ & 504 & 0.91 & 1.68 & 44 & $25 \times 22$ & 177\\
         & $\phi^2$\,Ori & 2.0260 & {\tt cam2} & $V$ & ND1 & 8.20 & $ 3.5 \times 6 $ & $3 \times 2$ & 504 & 0.91 & 1.68 & 26 & $25 \times 18$ & 179 \\
        \hline                      
        \noalign{\smallskip}
        4 & $\phi^2$\,Ori & 2.0362 & {\tt cam1} & CntH$\alpha$ & None & 100 & $ 3.5 \times 6 $ & $3 \times 2$ & 504 & 0.85 & 1.80 & 36 & $25 \times 22$ & 178\\
         & $\phi^2$\,Ori & 2.0362 & {\tt cam2} & NH$\alpha$ & None & 100 & $ 3.5 \times 6 $ & $3 \times 2$ & 504 & 0.85 & 1.80 & 40 & $22 \times 20$ & 169\\
        \hline                      
        \end{tabular}
        \tablefoot{MJD is the average modified julian date. ND indicates if a neutral density filter has been used; its transmission is listed in the Trans. column. DIT is the exposure time of the individual frames, and NDIT is the number of frames per exposure. Dither is the number of dithering positions ($\pm 14$\,pix), and Exp. the number of exposures per position.  $\theta$ is the visible seeing, and AM the airmass. $S$ is the Strehl ratio of the PSF images. $\theta_\mathrm{PSF}$ is the FWHM of the PSF images (Gaussian fit), and PA is the position angle of the major axis of the PSF elliptical core (north = $0^\circ$).}
\end{table*}

\begin{table}
        \caption{Photometry of Betelgeuse.}
        \centering          
        \label{phot}
        \begin{tabular}{lllc}
	\hline\hline
        \noalign{\smallskip}
	Filter & $\lambda_0$ & $\Delta \lambda$ & Flux \\
	 &  [nm] & [nm] & [$10^{-8}$\ W\,m$^{-2}$\,$\mu$m$^{-1}$] \\
        \noalign{\smallskip}
        \hline    
        \noalign{\smallskip}
         $V$ & 554 & 80.6 & $3.35 \pm 0.17$ \\
         CntH$\alpha$ & 644.9 & 4.1 & $6.29 \pm 0.31$ \\
         NH$\alpha$ & 656.34 & 0.97 & $6.23 \pm 0.31$ \\
         TiO717 & 716.8 & 19.7 & $4.70 \pm 0.23$ \\  
        \hline                      
        \end{tabular}
        \tablefoot{$\lambda_0$ is the central wavelength and $\Delta \lambda$ the bandwidth FWHM.}
\end{table}

\begin{figure*}[]
        \centering
	\includegraphics[width=4.5cm, page=1]{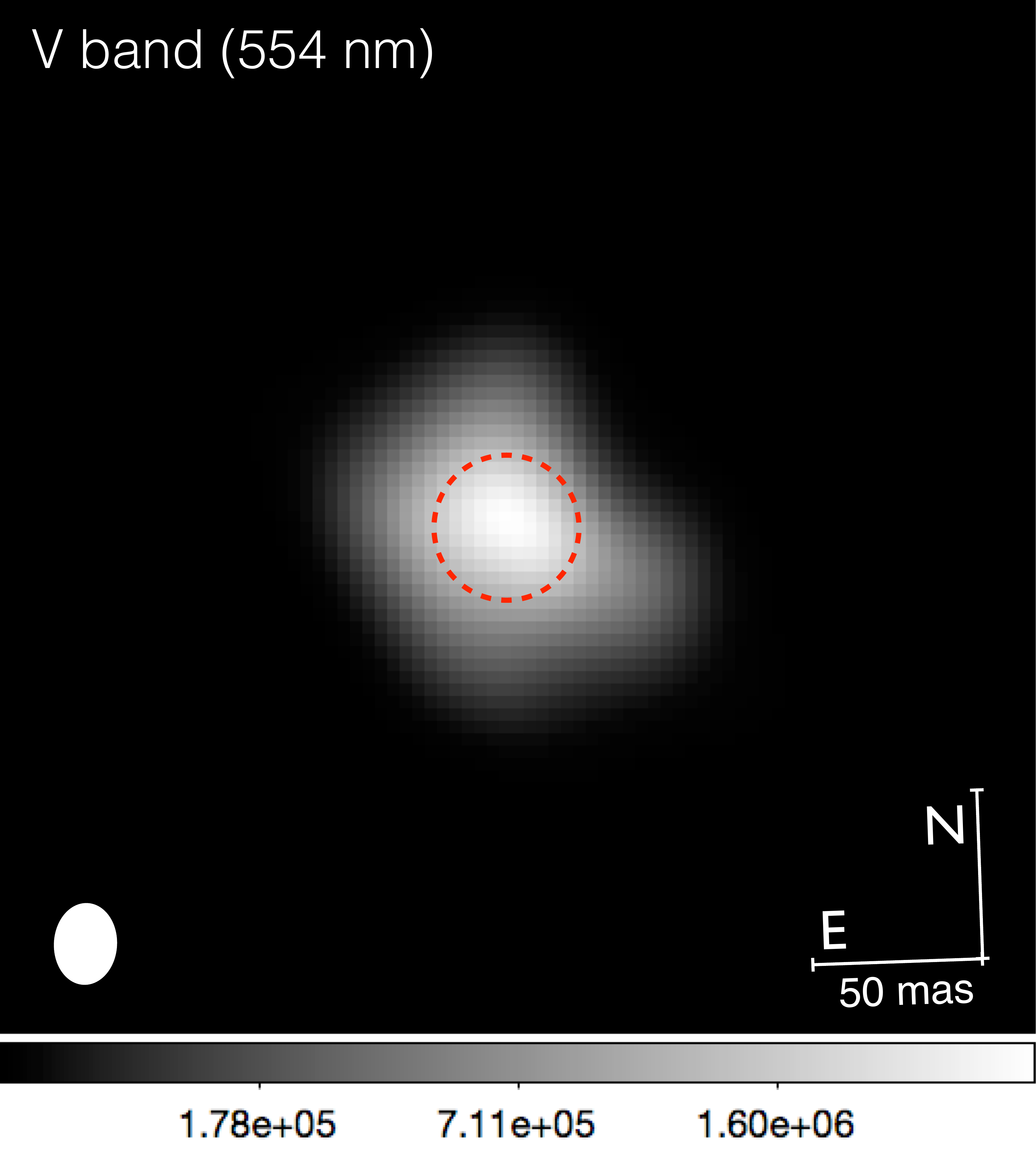}
	\includegraphics[width=4.5cm, page=2]{Figures/Figures-Betelgeuse-ZIMPOL.pdf}
	\includegraphics[width=4.5cm, page=3]{Figures/Figures-Betelgeuse-ZIMPOL.pdf}
	\includegraphics[width=4.5cm, page=4]{Figures/Figures-Betelgeuse-ZIMPOL.pdf}
	\includegraphics[width=4.5cm, page=9]{Figures/Figures-Betelgeuse-ZIMPOL.pdf}
	\includegraphics[width=4.5cm, page=10]{Figures/Figures-Betelgeuse-ZIMPOL.pdf}
	\includegraphics[width=4.5cm, page=11]{Figures/Figures-Betelgeuse-ZIMPOL.pdf}
	\includegraphics[width=4.5cm, page=12]{Figures/Figures-Betelgeuse-ZIMPOL.pdf}
        \caption{{\it Top row}: Intensity images $I$, with a square root gray scale from minimum to maximum intensity (in W\,m$^{-2}$\,$\mu$m$^{-1}$\ sr$^{-1}$). The dashed red circle represents the size of the photosphere in the near-infrared (R$_\star$). The beam size (FWHM of the PSF star images) is shown as a white ellipse in the lower left corner.
        {\it Bottom row:} Maps of the degree of linear polarization $p_L$, with a linear color scale from 0 to 12\% (identical for all maps). The field of view is $302 \times 302$\,mas, and all frames have been deconvolved using IRAF's {\tt lucy} algorithm (80 steps).
        \label{deconv80}}
\end{figure*}

\begin{figure*}[]
        \centering
	\includegraphics[width=4.5cm, page=5]{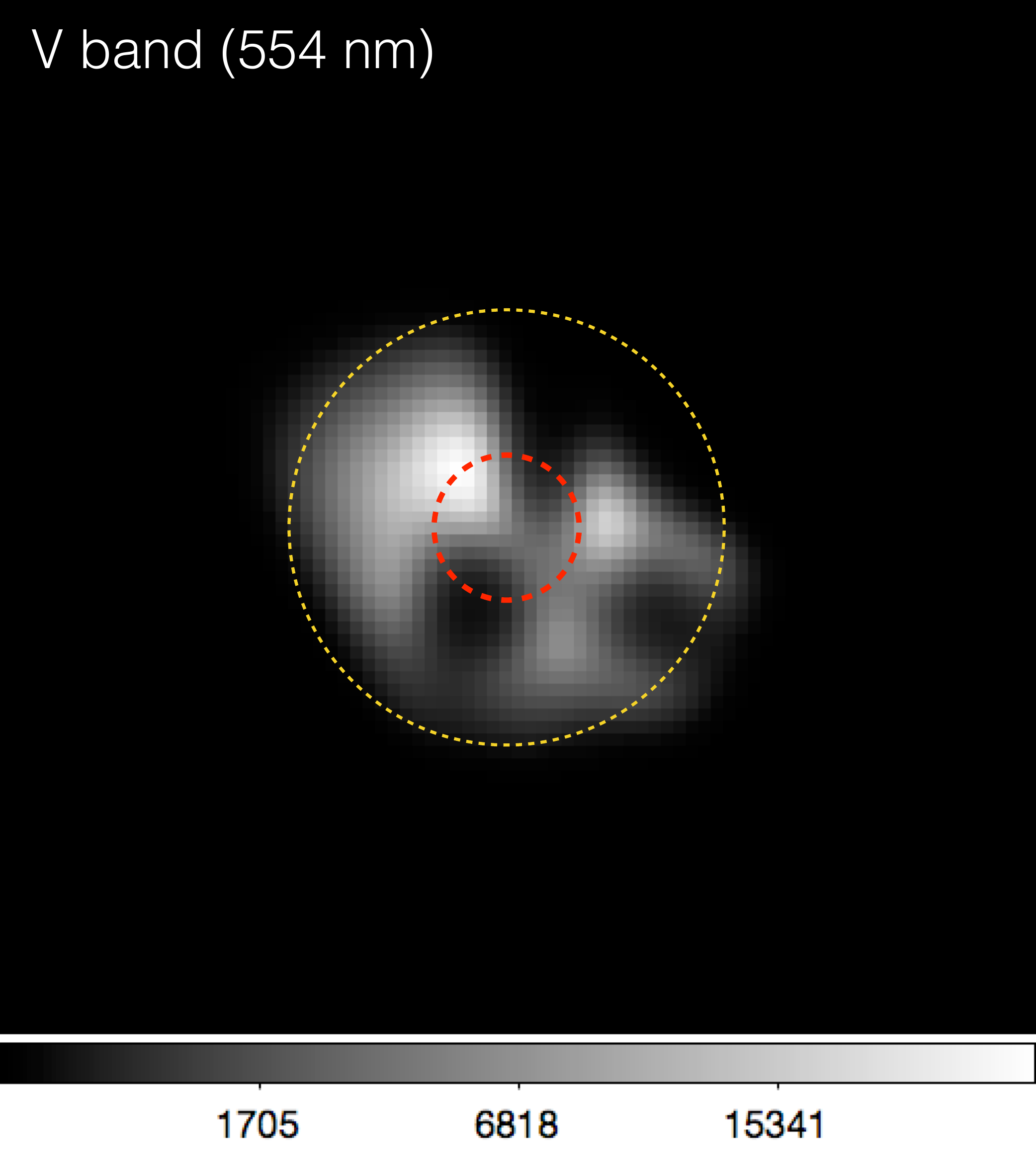}
	\includegraphics[width=4.5cm, page=6]{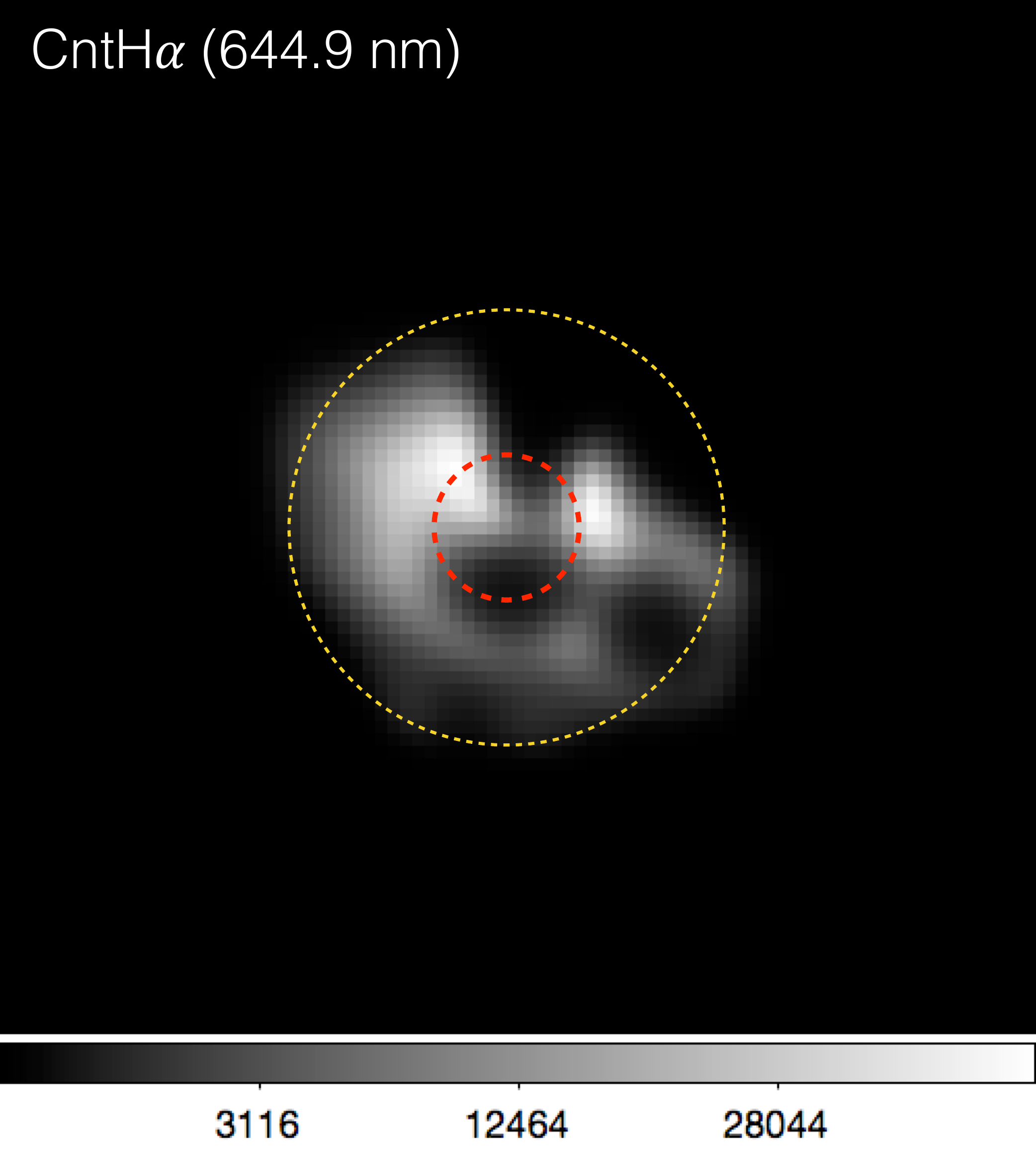}
	\includegraphics[width=4.5cm, page=7]{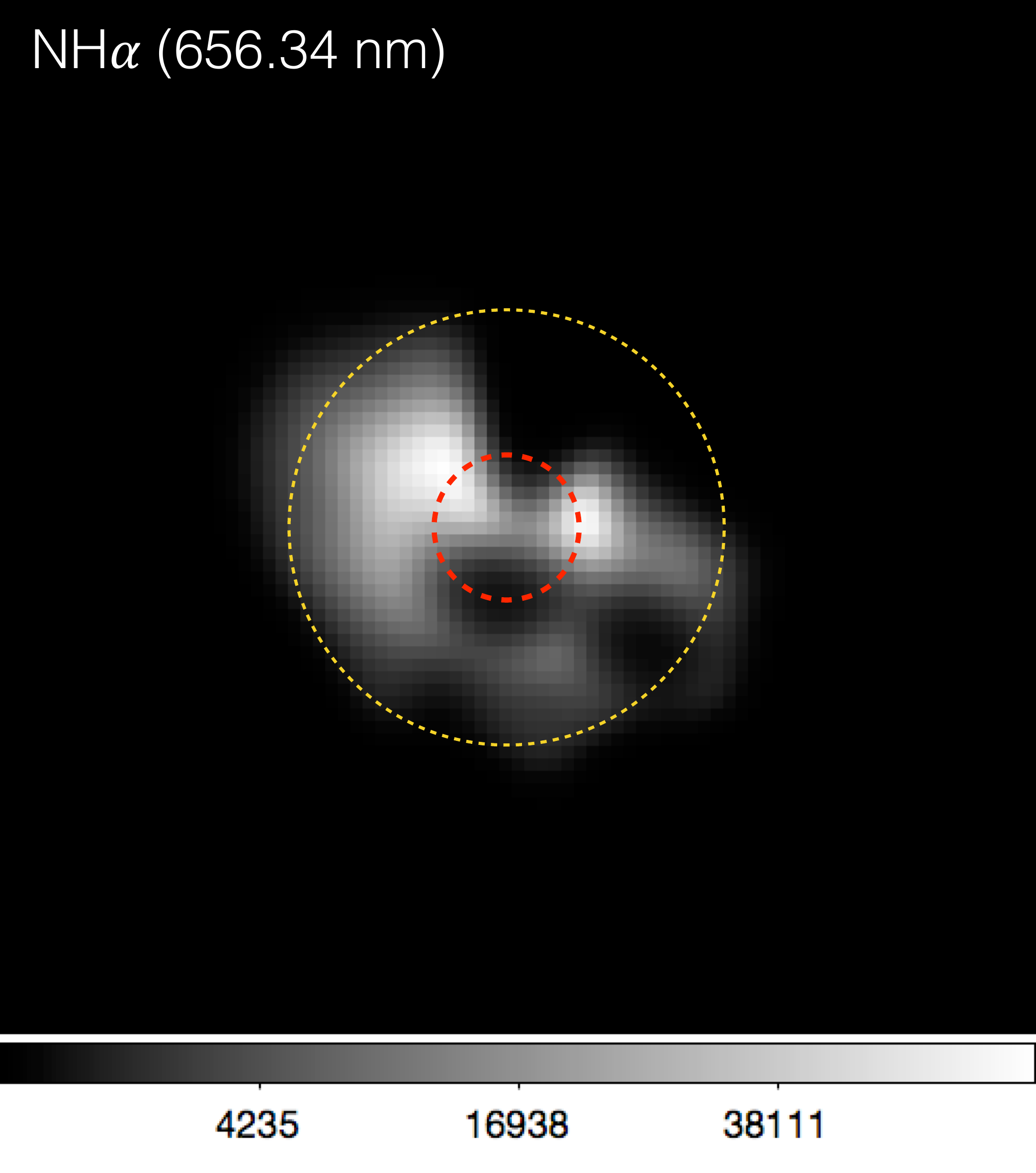}
	\includegraphics[width=4.5cm, page=8]{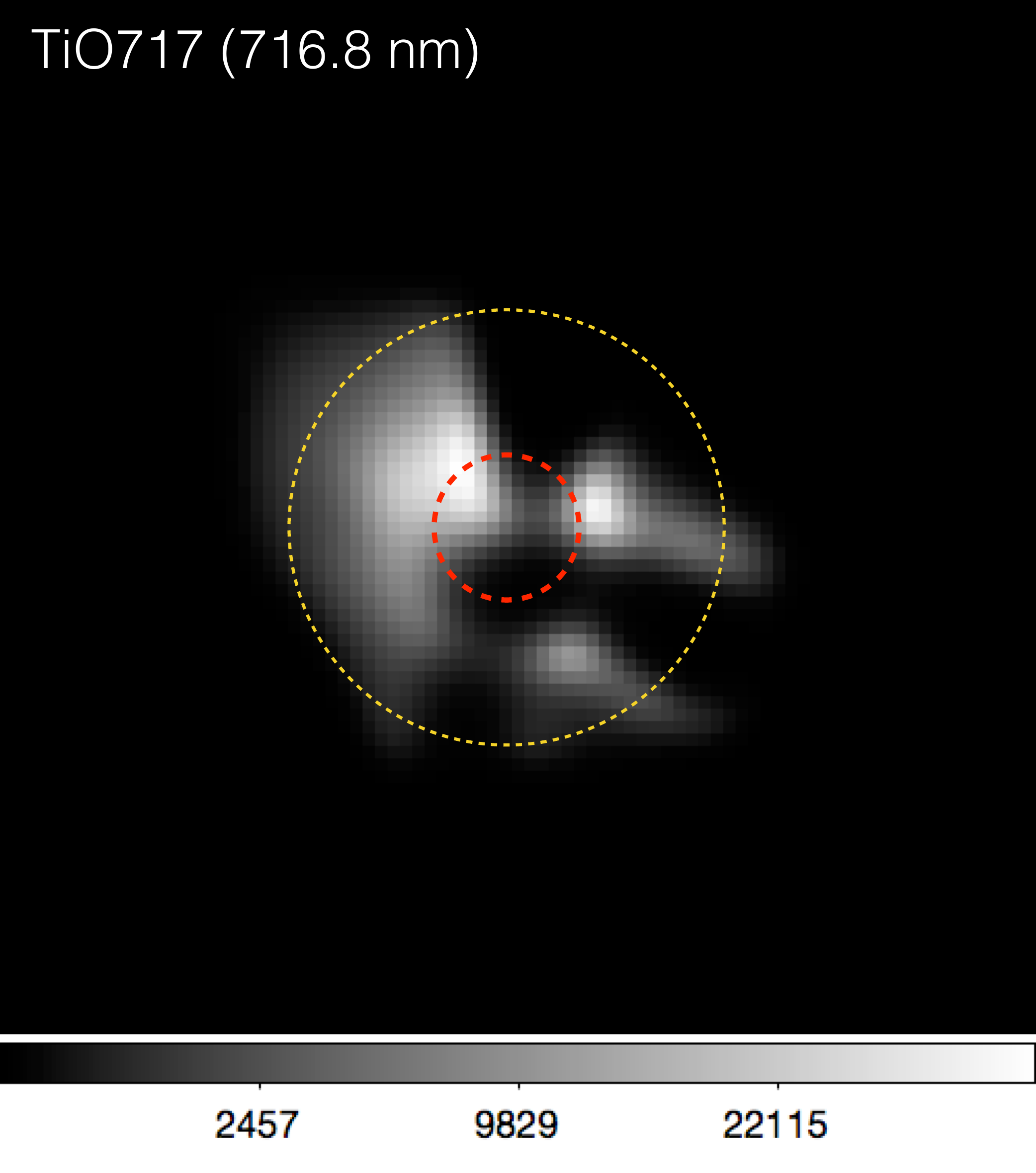}
	\includegraphics[width=4.5cm, page=16]{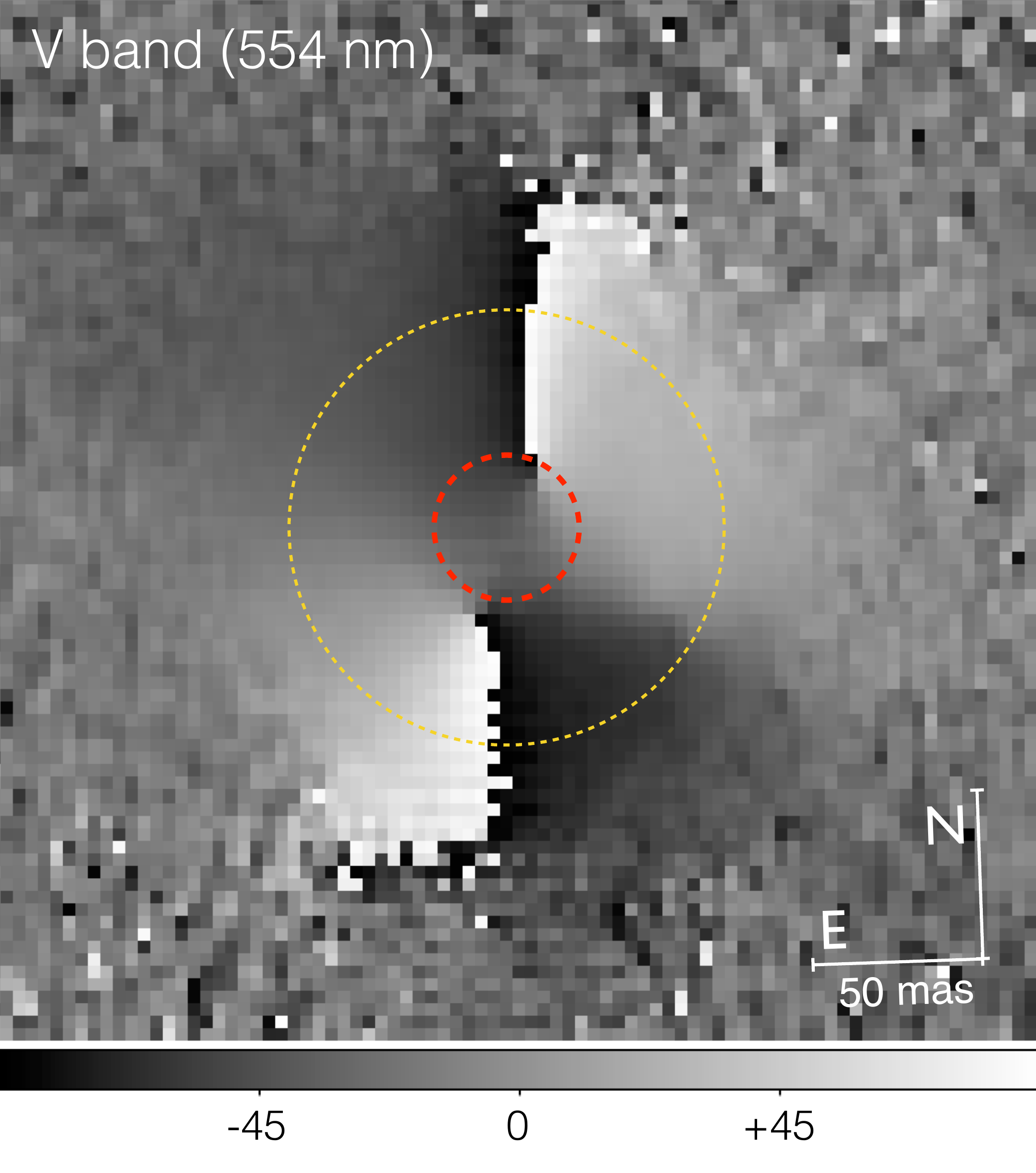}
	\includegraphics[width=4.5cm, page=17]{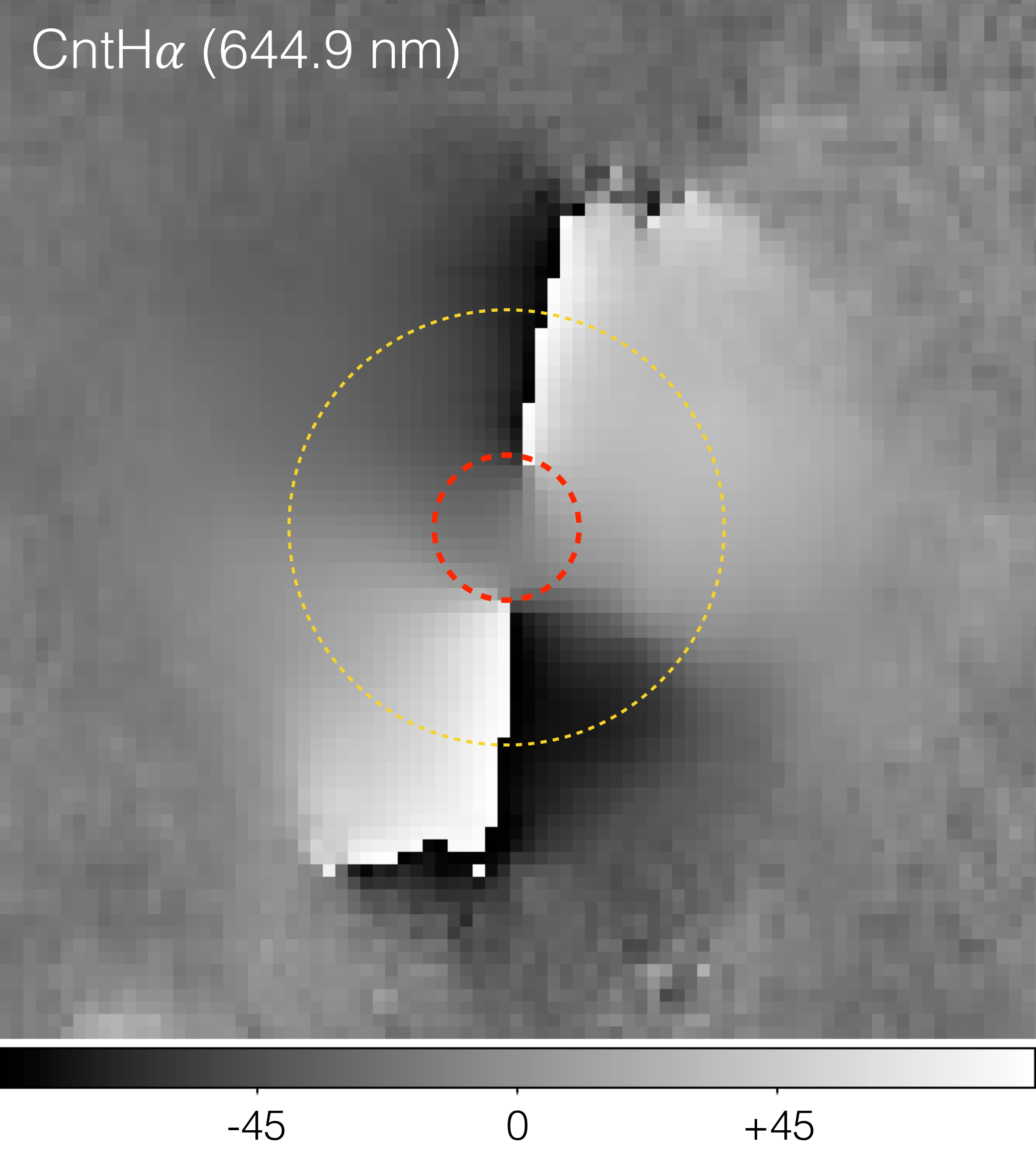}
	\includegraphics[width=4.5cm, page=18]{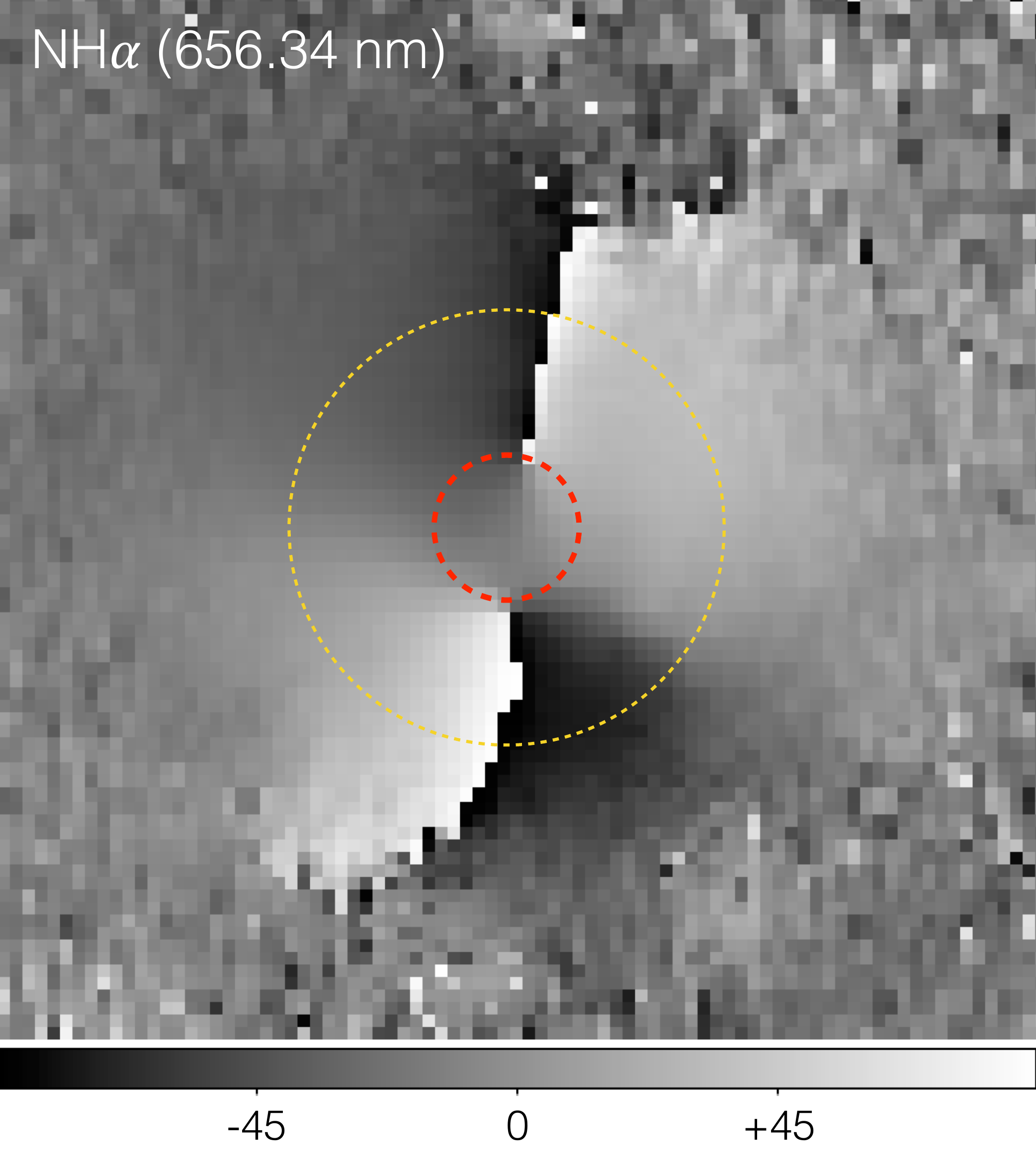}
	\includegraphics[width=4.5cm, page=19]{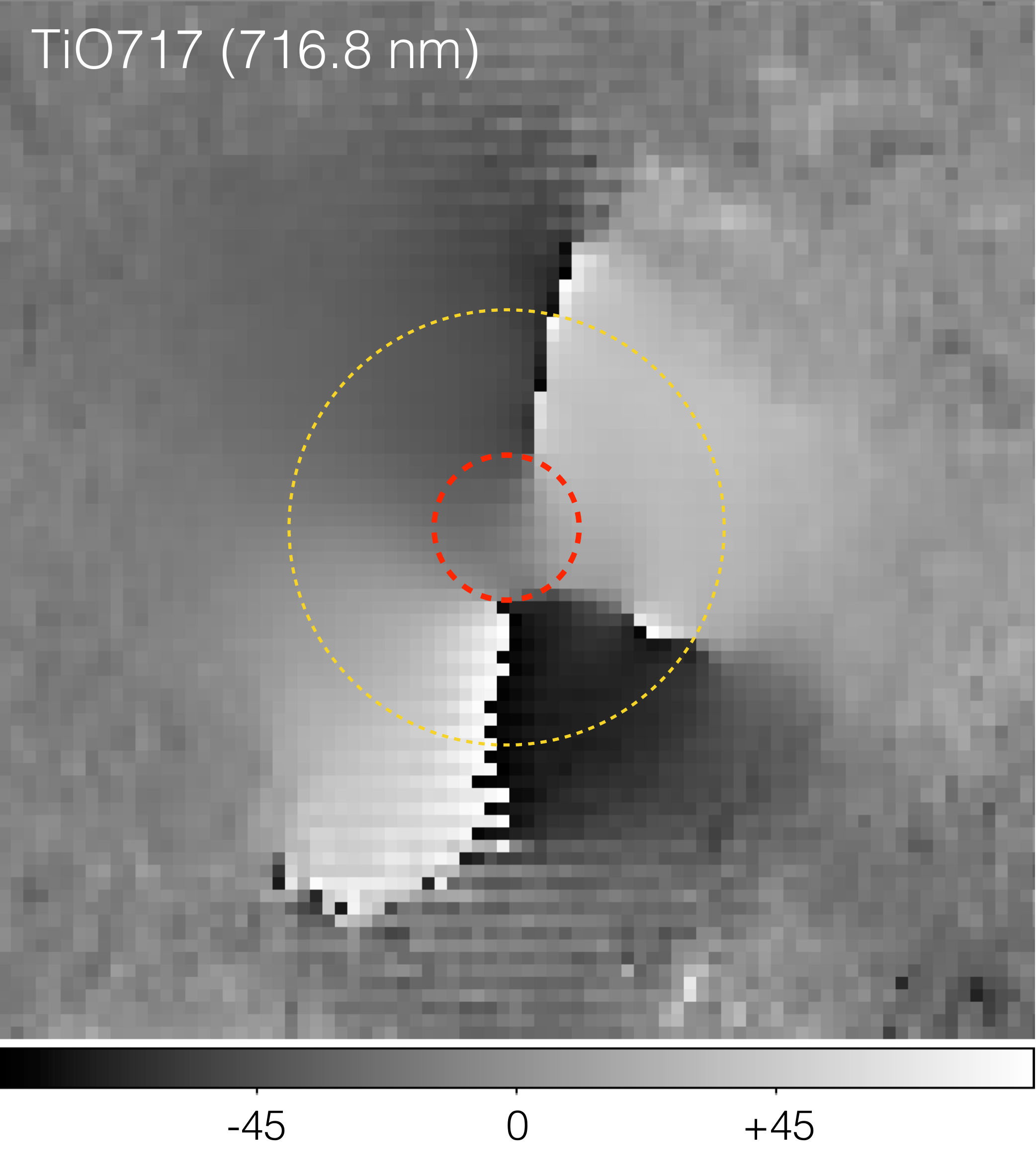}
        \caption{{\it Top row}: Polarized flux $P$ (square root scale from minimum to maximum in W\,m$^{-2}$\,$\mu$m$^{-1}$\ sr$^{-1}$).
        {\it Bottom row:} Polarization electric-vector position angle. The dashed red circle shows the photospheric size in the near-infrared (R$_\star$), and the yellow dashed circle has a radius of 3 R$_\star$.
        \label{polangle}}
\end{figure*}

We deconvolved the $I$, $P$ and $p_L$ images of Betelgeuse using the PSF total intensity maps $I$ as dirty beam and the Lucy-Richardson (L-R) algorithm implemented in IRAF.
We stopped the L-R deconvolution after 80 iterations, as the deconvolved images do not show a significant evolution for additional processing steps. The resulting $I$ and $p_L$ deconvolved frames are presented in Fig.~\ref{deconv80}. The maps of the polarized flux $P$ and of the polarization electric-vector position angle $\theta$ are shown in Fig.~\ref{polangle}.

\section{Analysis}\label{analysis}

\subsection{Photosphere and radial intensity profile\label{photosphere}}

The shape of the visible photosphere of Betelgeuse (Fig.~\ref{deconv80}, top row) is found to deviate noticeably from spherical symmetry. The extension of the central brighter part of the star image is comparable to the infrared photospheric size. This relatively circular central component appears surrounded by extended plumes, the most prominent extending to the southwest (it is also observed in the infrared by \citeads{2009A&A...504..115K}). We also notice less extended northern and southern plumes.
\begin{figure}[]
        \centering
        \includegraphics[width=8.0cm]{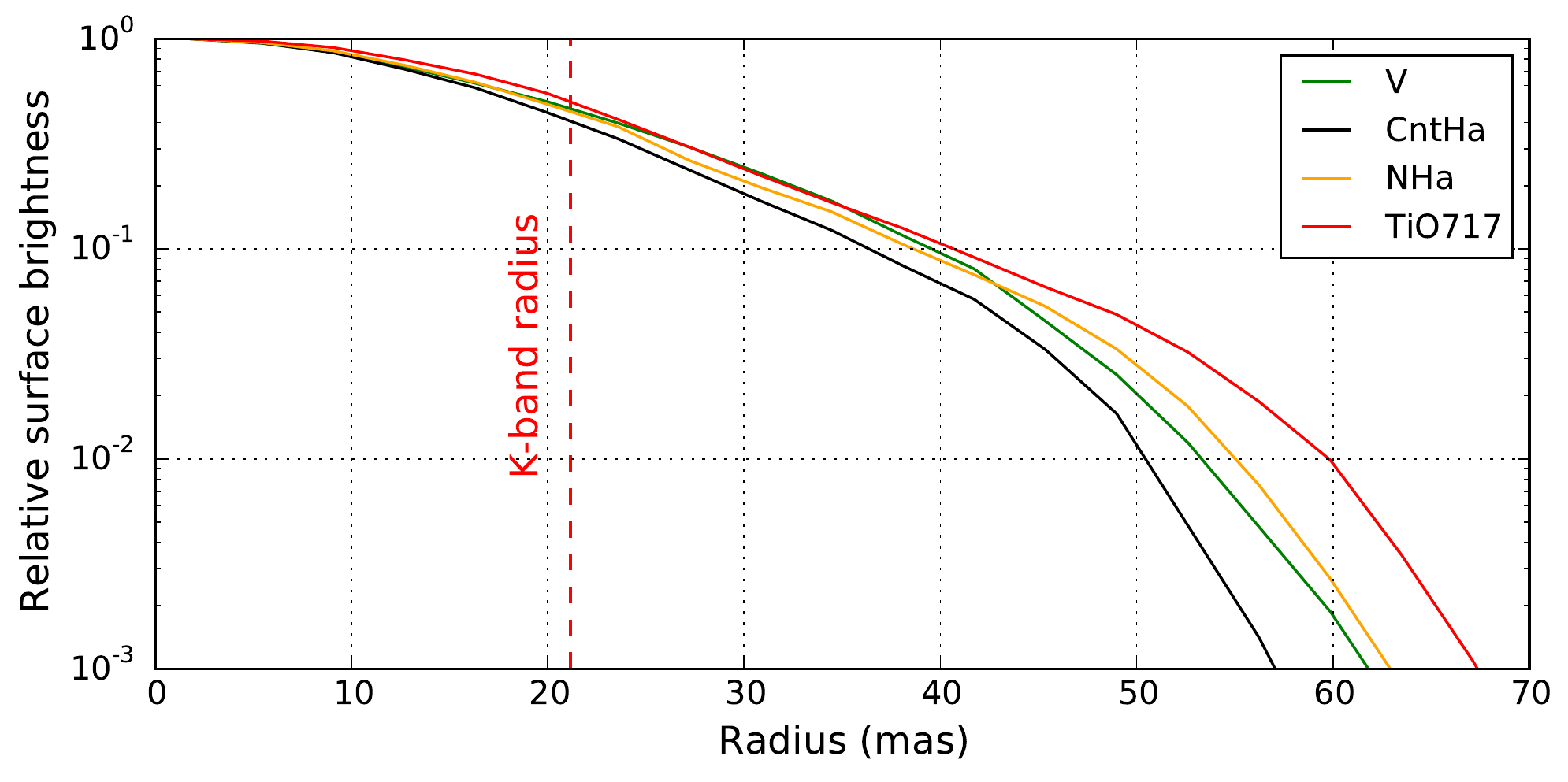}
        \caption{Radial median intensity profile of the deconvolved images of Betelgeuse.
        The $K$ band radius is indicated by a red dashed line.
        \label{profiles}}
\end{figure}
Figure~\ref{profiles} shows the radial median profiles of the intensity images of Betelgeuse in the four filters. The median angular diameters at half maximum are $\theta_\mathrm{50\%}(V) = 40.1$\,mas, $\theta_\mathrm{50\%}(\mathrm{CntH}\alpha) = 37.1$\,mas, $\theta_\mathrm{50\%}(\mathrm{NH}\alpha) = 39.3$\,mas and $\theta_\mathrm{50\%}(\mathrm{TiO717}) = 42.6$\,mas.
These sizes are consistent with the near-infrared photospheric (i.e. Rosseland) angular diameter.
%
The diameters at 1\% of the maximum are $\theta_\mathrm{1\%}(V) = 107$\,mas, $\theta_\mathrm{1\%}(\mathrm{CntH}\alpha) = 102$\,mas, $\theta_\mathrm{1\%}(\mathrm{NH}\alpha) = 111$\,mas and $\theta_\mathrm{1\%}(\mathrm{TiO717}) = 120$\,mas.

\subsection{H$\alpha$ envelope\label{halpha}}

\begin{figure}[]
        \centering
        \includegraphics[width=8.3cm]{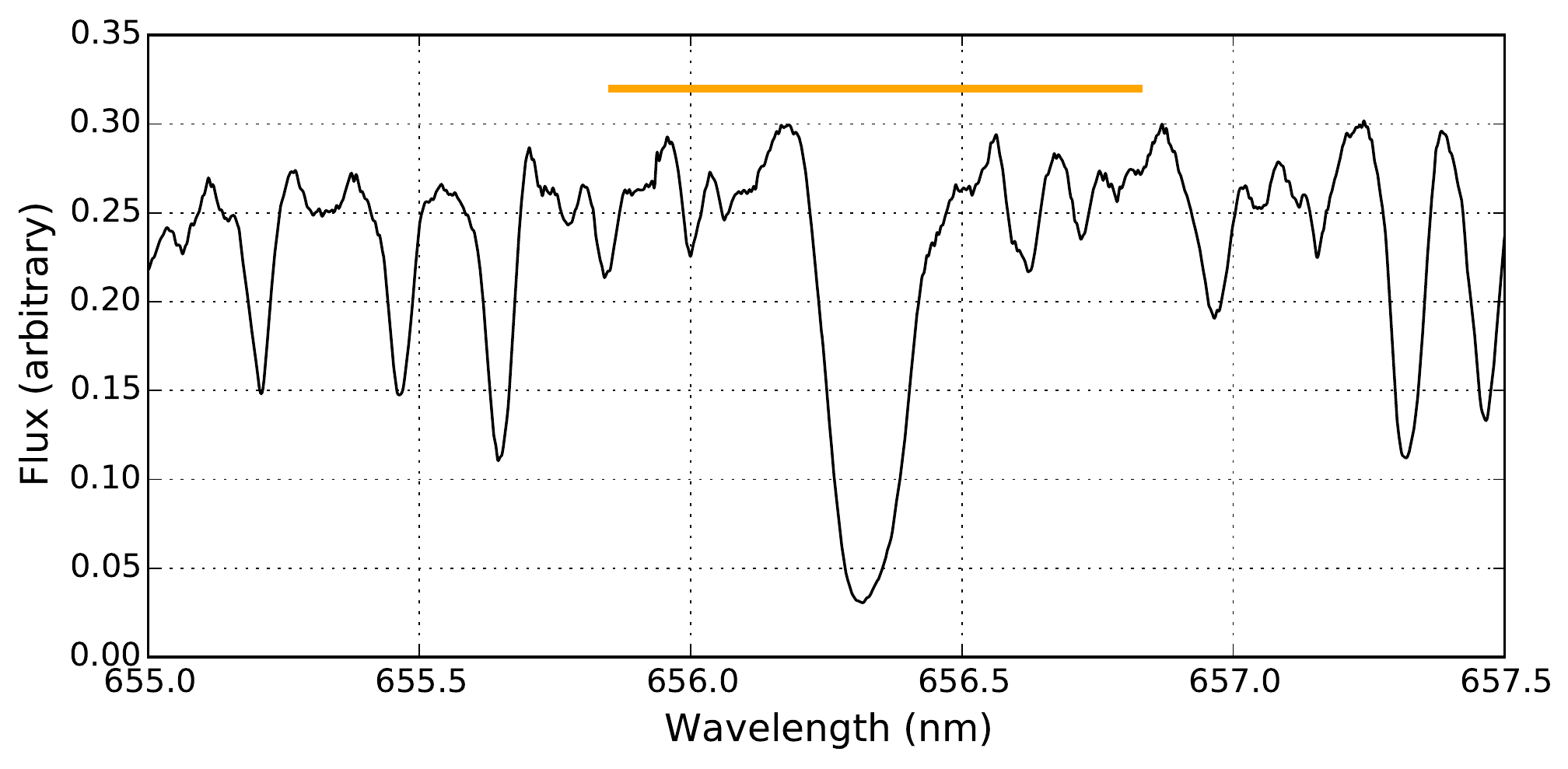}
        \caption{CORALIE spectrum of Betelgeuse around the H$\alpha$ line, with the NH$\alpha$ filter bandpass shown as an orange segment.
        \label{spectrumNHa}}
\end{figure}

\begin{figure}[]
        \centering
        \includegraphics[width=4.4cm, page=14]{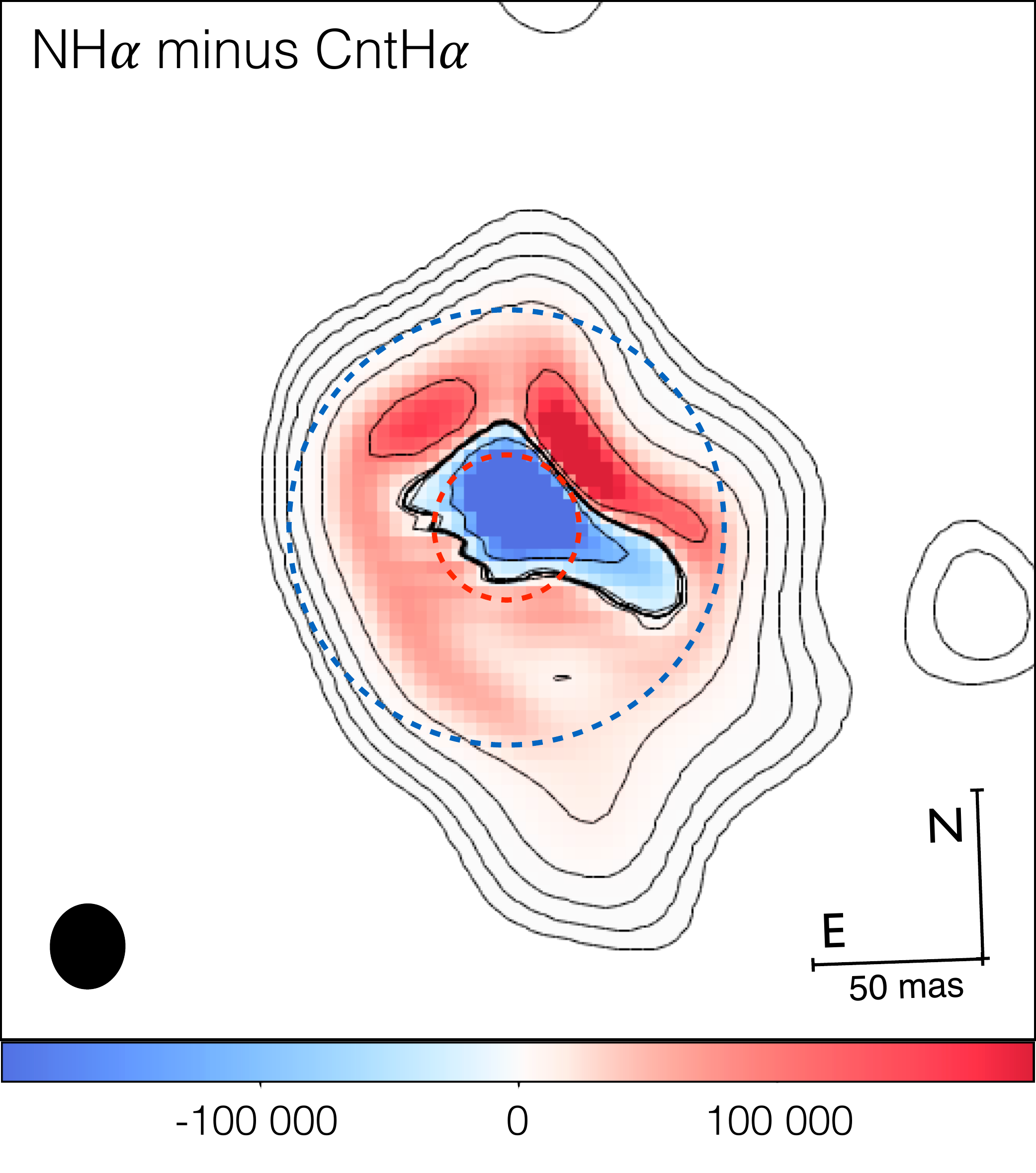}
	\includegraphics[width=4.4cm, page=15]{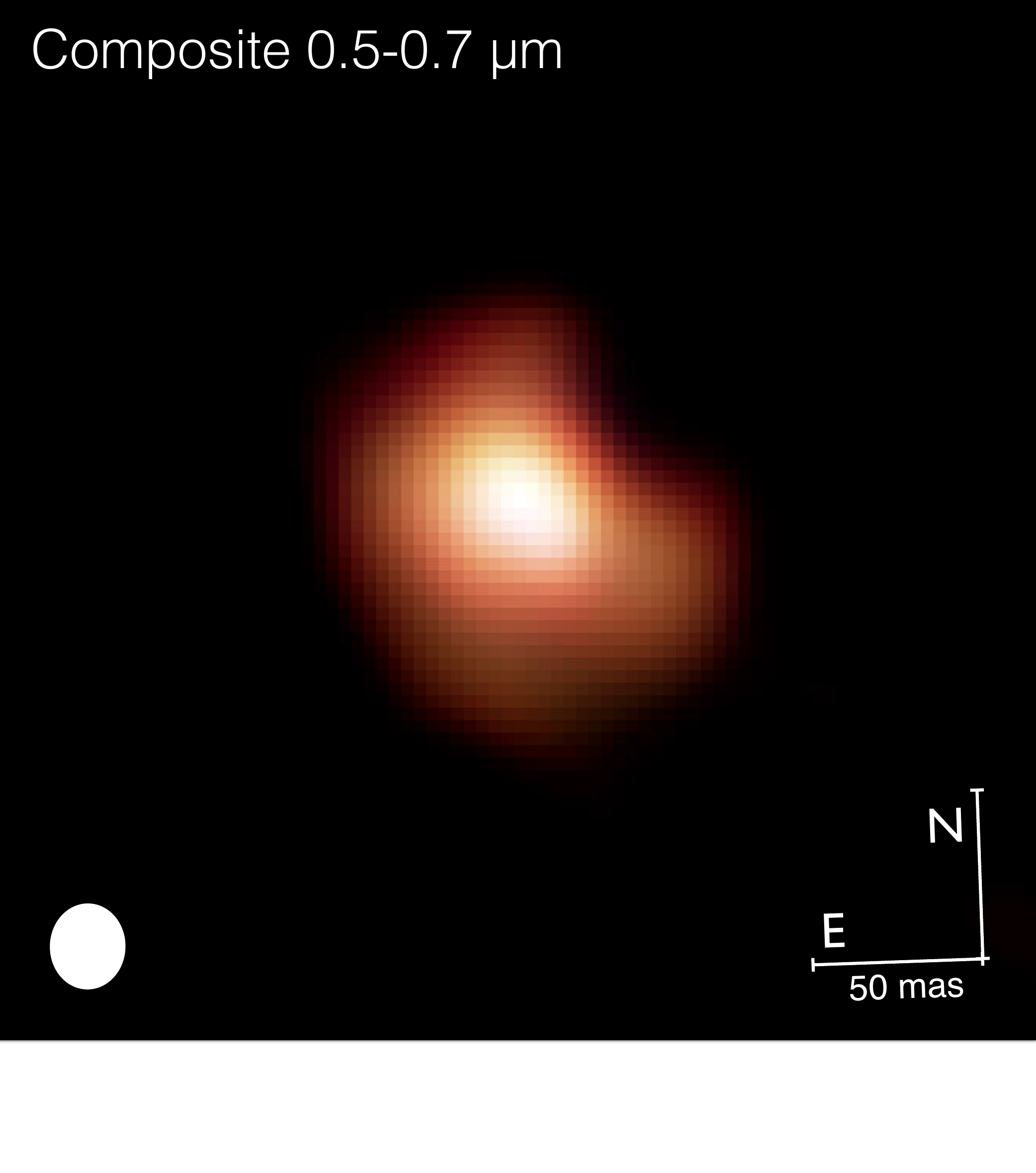} 
        \caption{{\it Left:} Map of the H$\alpha$ circumstellar emission of Betelgeuse. The color scale is linear in W\,m$^{-2}$\,$\mu$m$^{-1}$\ sr$^{-1}$. The contours are from 1 to $10^5$\ W\,m$^{-2}$\,$\mu$m$^{-1}$\ sr$^{-1}$ in absolute value, with a factor 10 spacing. The dashed red circle shows the infrared photosphere, and the dashed blue circle 3$\times$ its radius. The ellipse in the bottom left is the ZIMPOL beam. {\it Right:} Color composite of the $V$, CntH$\alpha$ and TiO717 images.
        \label{Halphasub}}
\end{figure}

The CntH$\alpha$ filter is designed to map the continuum flux distribution close to the H$\alpha$ wavelength. On Betelgeuse, the presence of a large number of electronic transitions in the visible results in a pseudo-continuum \citepads{2010A&A...515A..12C}.
The bandpass of the NH$\alpha$ filter is presented in Fig.~\ref{spectrumNHa} with a spectrum of Betelgeuse obtained on 22 April 2015 with the CORALIE instrument (\citeads{2001Msngr.105....1Q}, \citeads{2010A&A...511A..45S}). As the bandpass is relatively large, it includes a contribution from outside the H$\alpha$ line.
Once subtracted from the NH$\alpha$ filter image, the emission from the hydrogen envelope is revealed. Using the photometrically calibrated images, the result of this subtraction is shown in Fig.~\ref{Halphasub}. Both CntH$\alpha$ and NH$\alpha$ images are taken simultaneously using the two cameras of ZIMPOL, resulting in a perfect match of their PSFs.
The central part of the differential image has a negative intensity as the line is in absorption in front of the stellar disk. It is interesting to note that it is also in absorption in front of the main continuum plume extending to the southwest.
The H$\alpha$ emission is detected up to a radius of 5 R$_\star$. It appears inhomogeneous, and is essentially contained within 3\,R$_\star$ with a particularly strong emission around 2\,R$_\star$. A more extended H$\alpha$ plume is detected to the south of the star.

\subsection{Origin of the polarized flux and dust shell geometry\label{dustshell}}

\begin{figure}[]
        \centering
        \includegraphics[width=8.0cm]{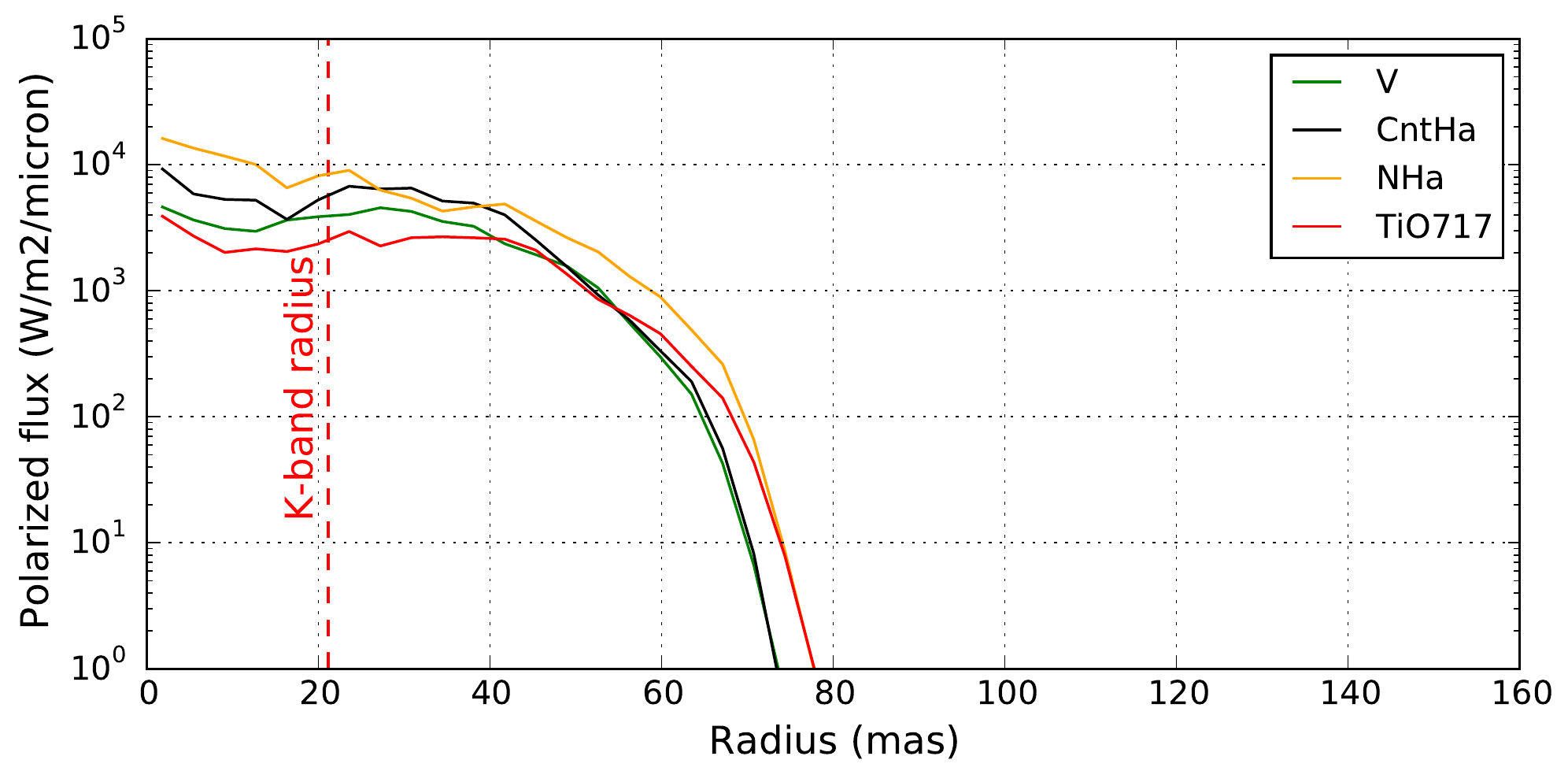}
        \includegraphics[width=8.0cm]{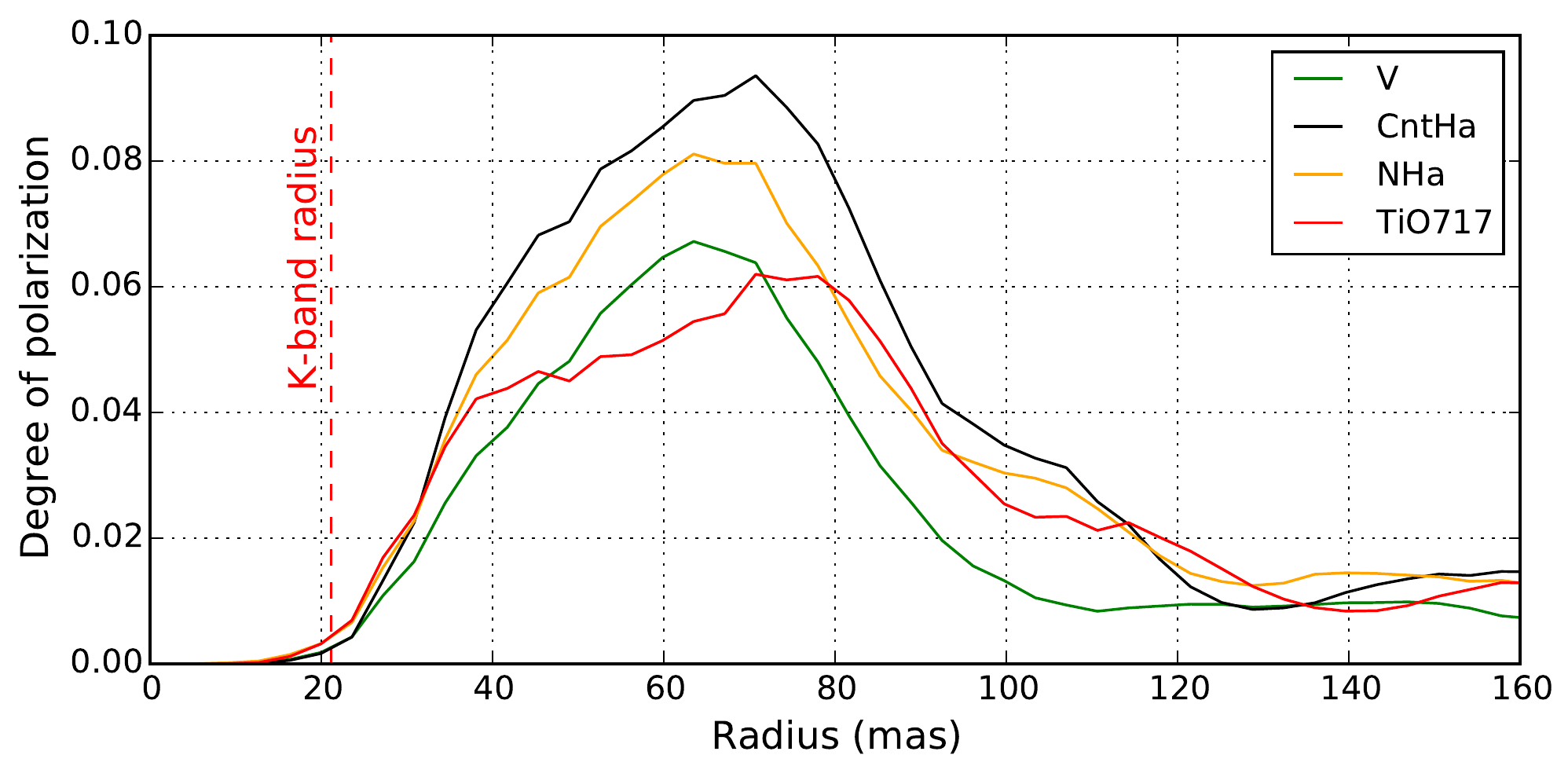}
        \caption{{\it Top panel:} Radial median profile of the polarized flux $P$. {\it Bottom panel:} Radial average profile of $p_L$ between azimuth 10$^\circ$ and 100$^\circ$. Both plots are based on the {\tt lucy} deconvolved maps (80\,steps).
        \label{poldeg-profiles}}
\end{figure}

A significant polarized flux $P$ is detected around Betelgeuse in all filters (Fig.~\ref{polangle}). It is asymmetric; its radial median profile is presented in Fig.~\ref{poldeg-profiles}.
This polarization could, in principle, be created by light scattering by dust, molecules, or even clumps of free electrons. We favor the dust hypothesis for several reasons. The interferometric observations by \citetads{2007A&A...474..599P} in the thermal infrared domain suggest that alumina is present within $1.5\,R_\star$. \citetads{2011A&A...531A.117K} also identified a compact unresolved emission at 20\,$\mu$m (their Fig. 6) that points to the presence of dust within 0.3$\arcsec$ of the star;
however, we cannot formally exclude a contribution to polarization from molecular scattering, as Betelgeuse is known to be surrounded by a molecular envelope.
The fact that we observe a significant $P$ flux superimposed over the photosphere indicates that multiple scattering may occur, as forward scattering is expected to produce little polarization.
The maps of the degree of linear polarization ($p_L$, Fig.~\ref{deconv80}) are sensitive to fainter contributions of the dust located at greater distances from the star where the background unpolarized flux contribution is lower. In typical Rayleigh conditions, high values of $p_L$ are created by $\approx 90^\circ$ scattering of the stellar light, while the scattered flux fraction is lower than in the forward scattering case.
Although the average value of $p_L$ over the field is small ($P/I = 0.49$\% in $V$, 1.28\% in CntH$\alpha$, 1.18\% in NH$\alpha$ and 0.97\% in TiO717), it reaches much higher values locally.
The $p_L$ maps show a prominent curved high-polarization region in the northeast quadrant of Betelgeuse at a radius of 2 to $3\,R_*$ and a fainter shell-like structure surrounding the star at a similar radius.
It is interesting to note that linear polarization in the $R$ and $I$ bands has been detected by \citetads{1994AAS...184.0507N} in Betelgeuse using the Wisconsin HPOL polarimeter (see also \citeads{1998AJ....116.2501U}). They indicate inhomogeneities at a position angle around $60^\circ$ (or $240^\circ$), consistent with the position angle of the partial shell that we observe.
The $p_L$ radial profile (Fig.~\ref{poldeg-profiles}) is similar for all filters, with a maximum around $3\,R_*$. The highest degree of linear polarization is observed in the CntH$\alpha$ filter and decreases for shorter and longer wavelengths. The slightly lower $p_L$ value in the NH$\alpha$ filter is likely caused by the presence of the unpolarized H$\alpha$ emission from the gaseous component of the CSE.
The maximum polarization occurs for scattering angles close to 90$^\circ$ for typical astrophysical dust (\citeads{2003ApJ...598.1017D}, see also \citeads{2014A&A...572A...7K}). We thus interpret this signal as the signature of a geometrically thin, spherical dust shell of inhomogeneous surface density.
The equilibrium temperature at radius $R$ is
${T_\mathrm{eq}} = {T_\mathrm{eff}} \sqrt[4]{1-a}\ \sqrt{{R_\star}/({2\, R})}$
where $a$ is the albedo. Assuming $a=0.1$ to $0.5$, we obtain $T_\mathrm{eq} = 1250$ to $1450$\,K for $T_\mathrm{eff} = 3641$\,K \citepads{2004A&A...418..675P}. This temperature is consistent with the sublimation of olivine particles \citepads{2011EP&S...63.1067K}, but the actual dust species, likely O-rich, remains uncertain. The 3\,$R_\star$ radius is larger than the radius measured by \citetads{2007A&A...474..599P} using MIDI (1.4\,$R_\star$), who suggested the presence of the highly refractory Al$_2$O$_3$ (alumina).
A dedicated radiative transfer modeling will be presented in a future publication, in particular with the aim of estimating the total dust mass.

\section{Conclusion}

We presented polarimetric images of Betelgeuse that resolve its visible photosphere and inner CSE.
The presence of gaseous plumes within 3\,$R_\star$ is consistent with the infrared observations by \citetads{2009A&A...504..115K}.
They indicate that the close environment of Betelgeuse is not spherically homogeneous, and that significantly different physical conditions (temperature, gas density) exist at the same radius from the star. Departure from spherical symmetry is also found in the dust envelope at large distances from the star (\citeads{2011A&A...531A.117K}, \citeads{2012A&A...548A.113D}) and it may play a fundamental role in the formation of molecules and dust grains.
The observed degree of linear polarization indicates the presence of light scattering dust at 3\,$R_\star$ that survives at $T \approx 1300$\,K.
The detected H$\alpha$ emission presents an inhomogeneous distribution, mostly around 2\,$R_\star$.
A speculative interpretation for the inhomogeneity of the dust layer is that the hot gaseous plumes may locally sublimate the dust, resulting in the observed irregular dust density in the shell.
We note that dust may be present closer to the photosphere, but its polarimetric signature is likely masked by the unpolarized emission from the gaseous CSE. 
The 3\,$R_\star$ radius emerges as an important interface between the internal gaseous CSE and the dust forming region. The presence of dust so close to the star may play an important role in the mass loss mechanism, for example, through radiative pressure on dust grains.
The dust shell identified between 0.5 and $1\arcsec$ by \citetads{2011A&A...531A.117K} in the thermal infrared could correspond to other types of oxygen-rich dust.
The environment of Betelgeuse thus appears as radially structured and generally departing from spherical symmetry.
Future observations of the time evolution of the inner envelope with AO and interferometry will bring valuable insight to its complex dynamics.

\begin{acknowledgements}
We are grateful to the SPHERE instrument team for the successful execution of the observations in difficult conditions.
We acknowledge financial support from the ``Programme National de Physique Stellaire'' (PNPS) of CNRS/INSU, France.
The CORALIE observations were made under program CNTAC CN2015A-6.
STR acknowledges partial support by NASA grant NNH09AK731.
AG acknowledges support from FONDECYT grant 3130361.
This research received the support of PHASE, the partnership between ONERA, Observatoire de Paris, CNRS and University Denis Diderot Paris 7.
PK and AG acknowledge support of the French-Chilean exchange program ECOS-Sud/CONICYT.
We acknowledge with thanks the variable star observations from the AAVSO International Database contributed by observers worldwide and used in this research.
This research made use of Astropy\footnote{Available at \url{http://www.astropy.org/}} \citepads{2013A&A...558A..33A}
the SIMBAD and VIZIER databases (CDS, Strasbourg, France) and NASA's Astrophysics Data System.
\end{acknowledgements}

\bibliographystyle{aa} 
\bibliography{biblioBetelgeuse}

\end{document}